\documentclass[aps,prd,10pt,notitlepage,nofootinbib,superscriptaddress]{revtex4-1}

\usepackage[utf8]{inputenc}
\usepackage{amsmath}
\usepackage{amssymb}
\usepackage{amsfonts}
\usepackage{newtxtext,newtxmath}
\usepackage{bm}
\usepackage{graphicx}
\usepackage[usenames,dvipsnames]{xcolor}
\usepackage{color}
\usepackage[colorlinks=true,linkcolor=blue,urlcolor=blue,citecolor=blue]{hyperref}

\usepackage{slashed}
\usepackage[english]{babel}
\usepackage{dcolumn}
\usepackage{blindtext}
\usepackage{epsfig}
\usepackage{pifont}
\usepackage{dsfont,mathrsfs}
\usepackage{cancel}
\usepackage{bigints}
\usepackage{accents}
\usepackage{soul}
\usepackage{multirow}
\usepackage{simpler-wick}
\usepackage{makecell}

\newcommand{\Ensembleaverage}[1]{\left\langle#1\right\rangle}

\newcommand{\ensembleaverage}[1]{\langle#1\rangle}

\newcommand{\FB}[1]{\left(#1\right)}
\newcommand{\SB}[1]{\left\{#1\right\}}
\newcommand{\TB}[1]{\left[#1\right]}

\newcommand{\fb}[1]{(#1)}

\newcommand{\nn}{\nonumber}
\newcommand{\del}{\partial}
\newcommand{\identity}{\mathds{1}}

\newcommand{\mcTc}{\mathcal{T}_C}

\newcommand{\scrL}{\mathscr{L}}

\newcommand{\sign}[1]{\text{sign}\left(#1\right)}
\newcommand{\munu}{{\mu\nu}}
\newcommand{\numu}{{\nu\mu}}
\newcommand{\alphabeta}{{\alpha\beta}}
\newcommand{\IM}{\text{Im}}
\newcommand{\RE}{\text{Re}}
\newcommand{\Tr}{\text{Tr}}

\newcommand{\utilde}{\widetilde{u}}

\newcommand{\psibar}{\overline{\psi}}
\newcommand{\Psibar}{\overline{\Psi}}

\newcommand{\pvec}{\bm{p}}
\newcommand{\xvec}{\bm{x}}
\newcommand{\qvec}{\bm{q}}
\newcommand{\kvec}{\bm{k}}

\newcommand{\wpr}{\omega_{\pvec}^r}
\newcommand{\wps}{\omega_{\pvec}^s}
\newcommand{\wkr}{\omega_{\kvec}^r}

\newcommand{\Wp}[1]{\omega_{\pvec}^{#1}}
\newcommand{\Wk}[1]{\omega_{\kvec}^{#1}}

\newcommand{\ppl}{p_+}
\newcommand{\pmi}{p_-}
\newcommand{\kpl}{k_+}
\newcommand{\kmi}{k_-}

\newcommand{\ran}{\text{ran}}

\newcommand{\tL}{\text{L}}
\newcommand{\tT}{\text{T}}

\newcommand{\tu}{\text{u}}
\newcommand{\td}{\text{d}}

\newcommand{\ut}[1]{\underaccent{\tilde}{#1}}

\begin{document}
\title{Electromagnetic spectral functions in hot and dense chirally imbalanced quark matter}

\author{Snigdha Ghosh}
\email{snigdha.physics@gmail.com}
\thanks{Corresponding Author}
\affiliation{Government General Degree College Kharagpur-II, Paschim Medinipur - 721149, West Bengal, India}

\author{Nilanjan Chaudhuri}
\email{sovon.nilanjan@gmail.com}
\affiliation{Variable Energy Cyclotron Centre, 1/AF Bidhannagar, Kolkata - 700064, India}
\affiliation{Homi Bhabha National Institute, Training School Complex, Anushaktinagar, Mumbai - 400085, India}

\author{Sourav Sarkar}
\email{sourav@vecc.gov.in}
\affiliation{Variable Energy Cyclotron Centre, 1/AF Bidhannagar, Kolkata - 700064, India}
\affiliation{Homi Bhabha National Institute, Training School Complex, Anushaktinagar, Mumbai - 400085, India}
\author{ Pradip Roy}
\email{pradipk.roy@saha.ac.in}
\affiliation{Saha Institute of Nuclear Physics, 1/AF Bidhannagar, Kolkata - 700064, India}
\affiliation{Homi Bhabha National Institute, Training School Complex, Anushaktinagar, Mumbai - 400085, India}

\begin{abstract}
The photon self-energy from chirally imbalanced quark matter is evaluated at finite temperature and density using the real time formulation of thermal field theory. The analytic structure is explored in detail exposing the cut structure which corresponds to a variety of physical scattering and decay processes in the medium and their thresholds. The mass of the quarks in the chiral symmetry broken phase are obtained from the gap equation of the Nambu--Jona-Lasinio model. It is found that, in presence of finite chiral chemical potential, the chiral condensate tends to get stronger at low temperature while the opposite is observed at high values of temperature. A continuous spectrum is obtained for the electromagnetic spectral function and this is purely a finite chiral chemical potential effect.
\end{abstract}

\maketitle
%
\section{INTRODUCTION}\label{sec.intro}
Relativistic collision of heavy ions aim to probe the QCD vacuum structure under extreme conditions of temperature and/or baryon density~\cite{Yagi:2005yb,Florkowski:2010zz,Heinz:2013th,Busza:2018rrf}. There exists in fact an infinite number of energy-degenerate different vacua of QCD characterized by an integer valued winding number and separated by a potential barrier~\cite{Shifman:1988zk}.  This is probed by non-trivial topological gauge configurations which interpolate between these vacua. 
At small values of temperature, transition between different vacua are dominated by instanton tunneling~\cite{Belavin:1975fg,tHooft:1976rip,tHooft:1976snw}, resulting in a lower transition rate. In Ref.~\cite{Schafer:1996wv} it is argued that, the existence of infrared instanton structure can provide a mechanism for the chiral phase transition. However, at temperatures comparable to the QCD scale, an abundant production of the QCD sphalerons, another kind of topological gluon configurations, is expected~\cite{Manton:1983nd,Klinkhamer:1984di} which can lead to the significant increase in the transition rate by crossing the barriers between different vacua~\cite{Kuzmin:1985mm,Arnold:1987mh,Khlebnikov:1988sr,Arnold:1987zg}. Interaction of the quarks with these topological gauge fields can change their helicities which in turn produces an imbalance between left and right handed quarks and thus lead to the breaking of the parity ($ P $) and charge-parity ($ CP $) symmetry by means of the axial anomaly of QCD~\cite{Adler:1969gk,Bell:1969ts}. As there is no direct $ P $ and $ CP $ violation in QCD, there can only be local domains with chirality imbalance which vanishes globally~\cite{McLerran:1990de,Moore:2010jd}. The existence of chiral imbalance or a difference in the number of right and left handed quarks would imply the existence of a chiral chemical potential (CCP).

Off-central collisions of heavy nuclei moving with velocities close to that of light can also generate a pulse of very strong magnetic field estimated to be $ \sim 10^{15} $ Tesla~\cite{Kharzeev:2007jp,Skokov:2009qp}. Consequently electromagnetic interactions driven by such high magnetic fields are as strong as QCD interactions. In this situation an asymmetry between the densities of left and right handed quarks will induce a separation of charges or an electric current being set up along the direction of the magnetic field. This is known as chiral magnetic effect (CME)~\cite{Fukushima:2008xe,Kharzeev:2007jp,Kharzeev:2009pj,Bali:2011qj}. Experimental detection of this transient effect will not only provide direct evidence for the formation of a chirally symmetric deconfined QCD matter with domains where $ P $ and $ CP $ symmetries are locally violated by QCD but also prove the existence of strong magnetic fields. Intense efforts have been going on to detect CME in heavy ion collision experiments at the RHIC at Brookhaven. Recently, the STAR Collaboration performed a blind analysis of a large data sample of approximately 3.8 billion isobar collisions which so far have yielded a null result~\cite{STAR:2021mii}. However, new methods for experimental determination of CME have been proposed~\cite{An:2021wof,Milton:2021wku}. Besides these, the chiral imbalance also leads to interesting physical phenomena. For example, chiral matter under rotation can generate currents along the vortical field, known as the chiral vortical effect which results in the chiral vortical waves~\cite{Vilenkin:1979ui,Vilenkin:1980fu,Son:2009tf,Jiang:2015cva}. In~\cite{Carignano:2018thu}, it is shown that, the damping rate depends on both the energy and the chirality of the fermion. The presence of chiral imbalance can also be responsible for a different propagation of the left and right circularly polarized transverse modes of the photon~\cite{Carignano:2018thu}, providing new contributions to the collisional energy loss of an energetic fermion~\cite{Carignano:2021mrn} and the fermion damping rate~\cite{Carignano:2019ivp}. Moreover, there exists unstable collective modes signaling the presence of a plasma instability in a system with an asymmetry between left- and right-handed chiral fermions which might provide a new mechanism for the strong and stable magnetic field of magnetars~\cite{Akamatsu:2013pjd,Matsumoto:2022lyb}.

Furthermore, it is encouraging to note that CME has been observed in condensed matter systems particularly in 3D Dirac as well as Weyl semimetals~\cite{Li:2014bha,Li:2016vlc,Kharzeev:2013ffa,Kharzeev:2015znc,Huang:2015oca,Landsteiner:2016led,Gorbar:2017lnp,Joyce:1997uy,Tashiro:2012mf}. In addition, there are also  important consequences of chiral imbalance on the phase structure of the strongly interacting matter. Recent studies~\cite{Ruggieri:2016lrn,Ruggieri:2016asg,Ruggieri:2020qtq} reveals that, in heavy ion collisions, chiral charge density reaches equilibrium shortly after the collision and the situation remains persistent for a larger period of time. Hence the study of chirally imbalanced matter continues to be of significant scientific relevance in the near future.

Correlation functions of local currents are among the primary theoretical tools which have been used to study the response of matter created in relativistic heavy ion collisions~\cite{Mallik:2016anp,Alam:1999sc, Sarkar:2012ty}. Of particular interest is the electromagnetic current correlation function which characterizes the response of the system to electromagnetic probes which have been extensively studied in the literature~\cite{McLerran:1984ay,Kajantie:1986dh,Weldon:1990iw,Alam:1996fd,Alam:1999sc,Rapp:1999ej,Aurenche:2000gf,Arnold:2001ms,Rapp:2009my,Chatterjee:2009rs}. The electromagnetic spectral function which is proportional to the imaginary part of the photon self-energy in the medium constitute the most important component in the definition of the emission rates of photons and dileptons. In this work, for the first time, we have derived the propagator of a massive fermion in presence of CCP. This propagator has been used to evaluate the photon self energy in chirally imbalanced matter using the real time formalism (RTF) of thermal field theory~\cite{Mallik:2016anp,Bellac:2011kqa,Kapusta:2006pm}. The analytic structure of this quantity is explored in detail exposing the cut structure which corresponds to a variety of physical scattering and decay processes in the medium and their thresholds. The ``strong'' interaction of quarks has been modeled using the 2-flavour Nambu--Jona-Lasionio (NJL) model~\cite{Nambu1,Nambu2} in the mean field approximation. This model has been extensively used to examine some of the nonperturbative properties of the QCD  at arbitrary values of temperature and baryon chemical potential (BCP) (see~\cite{Klevansky,Vogl,Buballa} for reviews), as well as in presence of a CCP~\cite{Farias:2016let,Ruggieri:2011xc,Fukushima:2010fe,Chao:2013qpa,Yu:2014sla,Yu:2015hym,Chaudhuri:2021lui}. The constituent quark mass at different external parameters, such as temperature, BCP and CCP, has been obtained by solving the corresponding gap equation. We find that in the presence of a CCP, the chiral condensate tends to get stronger at low temperature while it weakens at high values of temperature. These could be termed as `chiral catalysis' and `inverse chiral catalysis' respectively.

The article is organized as follows. In Sec.~\ref{sec.propagator}, we have shown the derivation of the real time fermion propagator at finite CCP. The propagator has then been used to evaluate the chiral condensate followed by the gap equation from the 2-flavour NJL model in Sec.\ref{sec.gap}. Sec.~\ref{sec.photon} is devoted for the calculation of the one-loop photon self energy in the medium with non-zero CCP. After that, the analytic structure of the thermal self energy function is analyzed in Sec.~\ref{sec.analytic}. Next in Sec.~\ref{sec.lorentz}, we have shown the Lorentz decomposition of the electromagnetic spectral function leading to various modes of photon propagation in the medium. In Sec.~\ref{sec.results}, we have shown and discussed the numerical results, and have finally summarized and concluded in Sec.~\ref{sec.summary}. Some calculational aspects are provided in the appendices.

\section{THE FERMION PROPAGATOR AT NON-ZERO CCP} \label{sec.propagator}
Let us consider the propagation of a spin-$\frac{1}{2}$ fermion of mass $M$ in a medium at zero temperature ($T=0$) with non-zero chemical potential $\mu$ and non-zero CCP $\mu_5 >0$. The system is described by the Lagrangian
	\begin{eqnarray}
	\scrL = \Psibar\FB{i\gamma^\mu \del_\mu + \mu \gamma^0 + \mu_5 \gamma^0 \gamma^5 -M}\Psi
	\label{eq.lag}
	\end{eqnarray}
where $\Psi$ denotes the fermion field. The corresponding coordinate space Dirac propagator $S(x,x')=S(x-x')$ satisfies~\cite{Kharzeev:2009pj}
\begin{eqnarray}
\FB{i\gamma^\mu \del_\mu + \mu \gamma^0 + \mu_5 \gamma^0 \gamma^5 -M} S(x-x') = -\delta^4(x-x').
\label{green.1}
\end{eqnarray}
To solve Eq.~\eqref{green.1}, let us introduce the Fourier transform of $ S(x-x') $ as
\begin{eqnarray}
S(x-x') = \int\frac{d^4p}{(2\pi)^4}e^{-ip\cdot(x-x')} S(p;M)
\label{fourier}
\end{eqnarray}
where, $S(p;M)$ is the momentum space Dirac propagator. Substituting \eqref{fourier} into Eq.~\eqref{green.1}, we obtain
\begin{eqnarray}
S(p;M) = \frac{-1}{\cancel{p}+ \mu \gamma^0 +\mu_5\gamma^0\gamma^5-M}. 
\label{prop.1}
\end{eqnarray}
Inverting the right hand side (RHS) of Eq.~\eqref{prop.1} is a bit involved; a long but straightforward calculation yields
\begin{eqnarray}
S(p;M) = \frac{-\mathscr{D}(p^0+\mu,\pvec;M)}{\widetilde{p}_+^2 \widetilde{p}_-^2 -2M^2 \widetilde{p}_+\cdot \widetilde{p}_- + M^4}
\label{prop.2}
\end{eqnarray} 
where $\widetilde{p}^\mu_\pm \equiv (p^0+ \mu \pm\mu_5,\pvec)$ and $\mathscr{D}(p;M)$ contains complicated Dirac structure as follows
\begin{eqnarray}
\mathscr{D}(p^0,\pvec;M) = \sum_{j \in \{\pm\}} \mathscr{P}_j \TB{ p_{-j}^2\cancel{p}_j - M^2 \cancel{p}_{-j} + M (p_j\cdot p_{-j}-M^2) 
	+ i M \sigma_\munu p_j^\mu p_{-j}^\nu  }
\label{D}
\end{eqnarray}
in which $\mathscr{P}_j = \frac{1}{2}(\identity+j\gamma^5)$ and $p^\mu_\pm \equiv (p^0 \pm\mu_5,\pvec)$.
The propagator in Eq.~\eqref{prop.2} can alternatively be expressed in the following form (which is easier to work with)
\begin{eqnarray}
S(p;M) = \mathscr{D}(p^0+\mu,\pvec;M) \sum_{r \in \{\pm\}} \frac{1}{4|\pvec|r\mu_5} \TB{ \frac{-1}{(p_0+\mu)^2 - (\wpr)^2 + i\epsilon} }
\label{prop.3}
\end{eqnarray}
where, $\wpr = \sqrt{(|\pvec|+r\mu_5)^2+M^2} >0$ and we have put an explicit $i\epsilon$ in the denominator following Feynman boundary condition. We note that, $r$ corresponds to the helicity of the propagating fermion~\cite{Kharzeev:2009pj}.

We now specify the fermion propagator at finite temperature ($T\ne0$) along with finite density ($\mu\ne0$) and finite CCP ($\mu_5\ne0$). For this, we will use the RTF of finite temperature field theory~\cite{Bellac:2011kqa,Mallik:2016anp,Kapusta:2006pm} where the thermal propagator assumes 2 × 2 matrix form. However, the knowledge of only $11$-component of this matrix is sufficient for our purpose, which is given by~\cite{Mallik:2016anp},
\begin{eqnarray}
S_{11}(p;M) = S(p;M) - \eta(p_0+\mu)\TB{S(p;M)-\gamma^0S^\dagger(p;M)\gamma^0}
\label{prop.4}
\end{eqnarray}
where $\eta(x)=\Theta(x)f_+(x) + \Theta(-x)f_-(-x)$ in which $f_\pm(x) = \TB{e^{(x\mp\mu)/T}+1}^{-1}$ is the Fermi-Dirac distribution function. Substituting Eq.~\eqref{prop.3} into Eq.~\eqref{prop.4} and simplifying, we get,
\begin{eqnarray}
S_{11}(p;M) = \mathscr{D}(p^0+\mu,\pvec;M) \sum_{r \in \{\pm\}} \frac{1}{4|\pvec|r\mu_5} 
\TB{ \frac{-1}{(p_0+\mu)^2 - (\wpr)^2 + i\epsilon} -\eta(p_0+\mu) 2\pi i \delta\FB{(p_0+\mu)^2-(\wpr)^2} }.
\label{prop.final}
\end{eqnarray}

It is to be noted that the fermion chemical potential $\mu$ has been put explicitly in the Lagrangian in Eq.~\eqref{eq.lag} for the calculation of thermo-dense propagator. In general, there exists two ways of incorporating fermion chemical potential in the real time formulation of thermal field theory: (i) explicitly through the Lagrangian or the equation of motion as done in this work in Eq.~\eqref{eq.lag}, or, (ii) through the density operator in the grand canonical (GC) ensemble~\cite{Niegawa:2002wj,Bellac:2011kqa}. Accordingly, there exists two different forms of the real time fermion propagator. However, in standard perturbative calculations the use of the two different forms of the propagator does not make any difference in the final result~\cite{Niegawa:2002wj}. A note on the introduction of fermion chemical potential in the RTF has been provided in Appendix~\ref{app.mu}.

\section{GAP EQUATION AND THE CONSTITUENT QUARK MASS FROM THE NJL MODEL} \label{sec.gap}
The Lagrangian (density) for the 2-flavour NJL model in presence of CCP reads,
\begin{eqnarray}
\scrL_\text{NJL} = \psibar \FB{i\gamma^\mu\del_\mu - m + \mu\gamma^0 + \mu_5 \gamma^0\gamma^5}\psi + G \TB{ (\psibar\psi)^2 + (\psibar i\gamma^5 \bm{\tau}\psi)^2}
\end{eqnarray}
where, $\psi = \begin{pmatrix}
\psi_\tu \\ \psi_\td
\end{pmatrix}$ 
is the quark isospin doublet, $m$ is the current quark mass and $G$ is the scalar channel coupling. Using the mean field approximation (MFA), the constituent quark mass $M$ can be calculated by solving the following gap equation:
\begin{eqnarray}
M = m - 2G \Ensembleaverage{\psibar\psi}
\label{gap}
\end{eqnarray}  
where, the chiral condensate $\Ensembleaverage{\psibar\psi}$ in MFA is given by 
\begin{eqnarray}
\Ensembleaverage{\psibar\psi} = N_c \sum_{f \in \{\tu,\td\} } \RE ~i \int\frac{d^4k}{(2\pi)^4}\Tr \TB{S_{11}(k;M)}.
\label{condensate.1}
\end{eqnarray}
Substituting Eq.~\eqref{prop.final} into Eq.~\eqref{condensate.1} and performing the $dk^0$ integral, we get after bit simplifications
\begin{eqnarray}
\Ensembleaverage{\psibar\psi} = -N_c M \sum_{f \in \{\tu,\td\} }\sum_{r \in \{\pm\}} \int \frac{d^3k}{(2\pi)^3} \frac{1}{\wkr} \TB{1-f_+(\wkr) -f_-(\wkr)}.
\label{condensate.2}
\end{eqnarray}
It is to be noted that, the temperature independent part of Eq.~\eqref{condensate.2} is ultraviolet (UV) divergent, which needs to be regularized. The NJL model, being non-renormalizable requires specific regularization scheme. In this work, we use the smooth three momentum cutoff prescription~\cite{Fukushima:2010fe} so that the gap equation becomes
\begin{eqnarray}
M = m + 2G N_c M \sum_{f \in \{\tu,\td\} }\sum_{r \in \{\pm\}} \int \frac{d^3k}{(2\pi)^3} \frac{1}{\wkr} \TB{  \sqrt{\frac{\Lambda^{20}}{\Lambda^{20} + |\kvec|^{20}}} -f_+(\wkr) -f_-(\wkr)}.
\label{gap.final}
\end{eqnarray} 
\begin{figure}[h]
	\begin{center}
		\includegraphics[angle=0,scale=0.35]{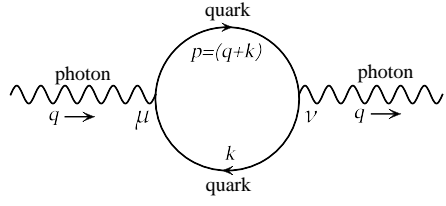} 
	\end{center}
	\caption{ Feynman diagram for one-loop photon self energy.}
	\label{fig.feynman}
\end{figure}

\section{PHOTON SELF ENERGY IN THE MEDIUM} \label{sec.photon}
The Lagrangian for photon-quark-antiquark electromagnetic interaction is $\mathscr{L}_\text{int}^\text{em} = J^\text{em}_\mu(x) A^\mu(x)$ where 
\begin{eqnarray}
J^\text{em}_\mu (x) = \sum_{f \in \{\tu,\td\} } e_f \psibar_f(x) \gamma_\mu \psi_f(x) 
\label{current}
\end{eqnarray}
is the conserved electromagnetic vector current that couples to the photon field $A^\mu(x)$. In Eq.~\eqref{current}, $e_f$ is the electronic charge of the quark flavour $f$. 
Using $\mathscr{L}_\text{int}^\text{em}$, 11-component of the one-loop real time photon self energy can be written as (applying Feynman rules to Fig.~\ref{fig.feynman})
\begin{eqnarray}
\Pi^\munu_{11}(q_0,\qvec) = -\sum_{f \in \{\tu,\td\} }e_f^2 N_c i \int\frac{d^4k}{(2\pi)^4} \Tr \TB{\gamma^\nu S_{11}(p=q+k;M)\gamma^\mu S_{11}(k;M)},
\label{photon.1}
\end{eqnarray}
where $N_c$ is the number of color. Substituting Eq.~\eqref{prop.final} into Eq.~\eqref{photon.1}, followed by performing the $dk^0$ integral, we get the imaginary part of the 11-component of in-medium photon self energy as
\begin{eqnarray}
\IM\Pi^\munu_{11} (q_0,\qvec) &=& N_c\sum_{f \in \{\tu,\td\} }e_f^2 \pi \int \frac{d^3k}{(2\pi)^3} \sum_{r \in \{\pm\}} \sum_{s \in \{\pm\}} 
\frac{1}{16rs\mu_5^2 |\pvec||\kvec|} \frac{1}{4\wkr\wps} \nn \\
&& \times \Big[ \mathcal{N}^\munu(k^0=-\wkr) \SB{1-f_-(\wkr)-f_+(\wps)+2f_-(\wkr)f_+(\wps)}\delta(q_0-\wkr-\wps) \nn \\
&& ~+~ \mathcal{N}^\munu(k^0=\wkr) \SB{1-f_+(\wkr)-f_-(\wps)+2f_+(\wkr)f_-(\wps)}\delta(q_0+\wkr+\wps) \nn \\
&& ~+~ \mathcal{N}^\munu(k^0=\wkr) \SB{-f_+(\wkr)-f_+(\wps)+2f_+(\wkr)f_+(\wps)}\delta(q_0+\wkr-\wps) \nn \\
&& ~+~ \mathcal{N}^\munu(k^0=-\wkr) \SB{-f_-(\wkr)-f_-(\wps)+2f_-(\wkr)f_-(\wps)}\delta(q_0-\wkr+\wps) \Big]
\label{impi.1}
\end{eqnarray}
where, 
\begin{eqnarray}
\mathcal{N}^\munu (k,q) = \Tr\TB{\gamma^\nu\mathscr{D}(p^0,\pvec;M)\gamma^\mu\mathscr{D}(k^0,\kvec;M)}.
\label{N.1}
\end{eqnarray}
On substituting Eq.~\eqref{D} into \eqref{N.1} and evaluating the Dirac trace, we will get the following expression of $\mathcal{N}^\munu (k,q)$ 
\begin{eqnarray}
\mathcal{N}^\munu(k,q) = 2g^\munu \big[ 2M^6 - M^4 \big\{2 (\ppl\cdot\pmi) + 2 (\kpl\cdot\kmi) + (\ppl\cdot\kpl) + (\pmi\cdot\kmi) \big\}
+ M^2 \big\{ (\ppl\cdot\kmi) (\pmi^2+\kpl^2) \nn \\
+ (\pmi\cdot\kpl) (\ppl^2 + \kmi^2) -2 (\ppl\cdot\kmi)(\pmi\cdot\kpl) + 2(\ppl\cdot\kpl)(\pmi\cdot\kmi) + 2(\ppl\cdot\pmi)(\kpl\cdot\kmi) \big\} \nn \\
-\ppl^2\kpl^2(\pmi\cdot\kmi) - \pmi^2\kmi^2(\ppl\cdot\kpl) \big]
+ 2\big[\ppl^\mu\kpl^\nu + \ppl^\nu\kpl^\mu\big] \big\{M^4-2M^2 (\pmi\cdot\kmi) + \pmi^2\kmi^2\big\} \nn \\
+2 \big[\pmi^\mu\kmi^\nu + \pmi^\nu \kmi^\mu\big] \big\{M^4-2M^2 (\ppl\cdot\kpl)+ \ppl^2\kpl^2 \big\}
-2M^2 \big[ \ppl^\mu\kmi^\nu + \ppl^\nu\kmi^\mu \big] (\pmi-\kpl)^2 \nn \\
-2M^2 \big[ \pmi^\mu\kpl^\nu + \pmi^\nu \kpl^\mu \big] (\ppl-\kmi)^2 
- 2i \varepsilon^{\mu\nu\alpha\beta} \big[ 2 p_{+\alpha}p_{-\beta} M^2 \SB{M^2-(\kpl\cdot\kmi)} \nn \\ 
- 2k_{+\alpha}k_{-\beta}M^2\SB{M^2-(\ppl\cdot\pmi)} + p_{+\alpha}k_{+\beta} \fb{M^4-\pmi^2\kmi^2} 
- p_{-\alpha}k_{-\beta} \fb{M^4-\ppl^2\kpl^2} \nn \\
+ M^2 p_{+\alpha}k_{-\beta} \fb{\pmi^2-\kpl^2} - M^2 p_{-\alpha}k_{+\beta} \fb{\ppl^2-\kmi^2}
\big]
\label{N.2}
\end{eqnarray}
in which we have used the convention $\varepsilon^{0123}=1$ for the four-dimensional Levi-Civita symbol. It is worth mentioning that, one of the integrations $d(\cos\theta)$ of Eq.~\eqref{impi.1} can analytically be performed using the Dirac delta functions present in the integrand.

\section{ANALYTIC STRUCTURE OF THE SELF ENERGY} \label{sec.analytic}
The imaginary part of the photon self energy at non-zero CCP in Eq.~\eqref{impi.1} contains sixteen Dirac delta functions. They give rise to branch cuts of the thermal self energy function in the complex $q_0$ plane. The terms with $\delta(q_0-\wkr-\wps)$ and $\delta(q_0+\wkr+\wps)$ are called Unitary-I and Unitary-II cuts respectively. On the other hand, the terms with $\delta(q_0+\wkr-\wps)$ and $\delta(q_0-\wkr+\wps)$ are respectively called Landau-I and Landau-II cuts. Each of the Unitary and Landau cut again consists of four sub-cuts corresponding to different helicities $ (r,s) $. These different cuts correspond to different physical processes like decay and scattering (absorption/emission). For example, the Unitary-I cuts correspond to the decay of a virtual photon having positive energy to real quark-antiquark pair (and the time reversed process). The Landau cuts correspond to the absorption (emission) processes in which a real quark/antiquark in the thermal medium absorbs (emits) a virtual photon.

Each of the sixteen Dirac delta functions in Eq.~\eqref{impi.1} are non-vanishing at different respective kinematic regions. To see this, we write 
$\delta(q_0\mp\wkr\mp\wps) = \delta(q_0\mp U^{rs})$ and $\delta(q_0\pm\wkr\mp\wps) = \delta(q_0\mp L^{rs})$
where,
\begin{eqnarray}
U^{rs} = U^{rs}(|\kvec|,x=\cos\theta; M,|\vec{q}|,\mu_5) &=& \wkr+\wps \label{U} \\
L^{rs} = L^{rs}(|\kvec|,x=\cos\theta; M,|\vec{q}|,\mu_5) &=& -\wkr+\wps  \label{L}
\end{eqnarray}
in which $\theta$ is the angle between $\qvec$ and $\kvec$. Both the functions $U^{rs}(|\kvec|,x)$ and $L^{rs}(|\kvec|,x)$ are defined in the domain $|\kvec| \in [0,\infty)$ and $x \in [-1,1]$. Therefore, the Unitary cuts, $\delta(q_0\mp\wkr\mp\wps)$ will be non vanishing iff 
\begin{eqnarray}
\pm q_0 \in \ran(U^{rs})
\label{unitary.1}
\end{eqnarray}
where, $\ran(f)$ denotes the range (or co-domain) of the function $f$. Eq.~\eqref{unitary.1} ensures that the spike of the Dirac delta function lies in the domain of integration in Eq.~\eqref{impi.1}. In a similar fashion, for the Landau cuts, $\delta(q_0\pm\wkr\mp\wps)$ will be non vanishing iff 
\begin{eqnarray}
\pm q_0 \in \ran(L^{rs}).
\label{landau.1}
\end{eqnarray}

For different values of $(r,s)$, the ranges $\ran(U^{rs})$ and $\ran(U^{rs})$ are provided in Appendix~\ref{app.domain} (see in Eqs.~\eqref{tab.range.u} and \eqref{tab.range.l}). Using Eqs.~\eqref{unitary.1} and \eqref{landau.1}, we can now find the kinematic regions for the sixteen Dirac delta functions, which are listed below
in Eqs.~\eqref{tab.kin.u} and \eqref{tab.kin.l}): 
\begin{eqnarray}
\begin{tabular}{|c|c|}
\hline 
Dirac Delta Function & \makecell[c]{ ~ \\ Kinematic Regions \\ ~ } \\
\hline \hline
$\delta(q_0\mp\Wk{+}\mp\Wp{+})$ & \makecell[c]{ ~ \\ $ \sqrt{(|\qvec|+2\mu_5)^2+4M^2} \le \pm q_0 < \infty$ \\ ~ }\\
\hline
$\delta(q_0\mp\Wk{+}\mp\Wp{-})$ & \makecell[c]{ ~ \\ $ \sqrt{\mu_5^2+M^2} + \sqrt{(|\qvec|-\mu_5)^2+M^2}  \le \pm q_0 < \infty $ for $ |\qvec| < 2\mu_5 $, \\
	$ \sqrt{\qvec^2+4M^2}  \le \pm q_0 < \infty $  for $ |\qvec| \ge 2\mu_5 $ \\ ~ } \\
\hline 
$\delta(q_0\mp\Wk{-}\mp\Wp{+})$ & \makecell[c]{~ \\ $ \frac{1}{2}\sqrt{(|\qvec|-2\mu_5)^2+4M^2} + \frac{1}{2}\sqrt{(|\qvec|+2\mu_5)^2+4M^2}  \le \pm q_0 < \infty $ for $ |\qvec| < \mu_5 $, \\
	$ \frac{1}{2}\sqrt{\qvec^2+4M^2} + \frac{1}{2}\sqrt{(||\qvec|-2\mu_5|+2\mu_5)^2+4M^2}  \le \pm q_0 < \infty$  for $ |\qvec| \ge \mu_5 $ \\ ~ } \\
\hline
$\delta(q_0\mp\Wk{-}\mp\Wp{-})$ & \makecell[c]{ ~ \\ $ 2M  \le \pm q_0 < \infty $ for $ |\qvec| < 2\mu_5 $, \\
	$  \sqrt{(|\qvec|-2\mu_5)^2+4M^2}  \le \pm q_0 < \infty$  for $ |\qvec| \ge 2\mu_5 $ \\ ~} \\	
\hline 
\end{tabular}
\label{tab.kin.u}~,
\end{eqnarray}
\begin{eqnarray}
\begin{tabular}{|c|c|}
\hline 
Dirac Delta Function & \makecell[c]{ ~ \\ Kinematic Regions \\ ~ } \\
\hline \hline
$\delta(q_0\pm\Wk{+}\mp\Wp{+})$ &  \makecell[c]{~ \\ $ -|\qvec| \le  q_0 \le |\qvec| $ \\ ~ } \\
\hline
$\delta(q_0\pm\Wk{+}\mp\Wp{-})$ & \makecell[c]{ ~ \\ $ -|\vec{q}|-2\mu_5 \le \pm q_0 \le -\sqrt{\mu_5^2+M^2} + \sqrt{(|\qvec|-\mu_5)^2+M^2} $ for $ |\qvec| < 2\mu_5 $, \\
	$ -|\vec{q}|-2\mu_5 \le \pm q_0 \le |\vec{q}|-2\mu_5 $  for $ |\qvec| \ge 2\mu_5 $ \\ ~ } \\
\hline 
$\delta(q_0\pm\Wk{-}\mp\Wp{+})$ & \makecell[c]{ ~ \\ $ -\frac{1}{2}\sqrt{(|\qvec|-2\mu_5)^2+4M^2} + \frac{1}{2}\sqrt{(|\qvec|+2\mu_5)^2+4M^2} \le \pm q_0 \le |\vec{q}|+2\mu_5 $ for $ |\qvec| < \mu_5 $, \\
	$ -\frac{1}{2}\sqrt{\qvec^2+4M^2} + \frac{1}{2}\sqrt{(|\qvec|-4\mu_5)^2+4M^2} \le \pm q_0 \le |\vec{q}|+2\mu_5 $  for $ \mu_5 \le |\qvec| < 2\mu_5 $, \\
	$ -|\vec{q}|+2\mu_5 \le \pm q_0 \le |\vec{q}|+2\mu_5 $  for $ |\qvec| \ge 2\mu_5 $ \\ ~ } \\
\hline
$\delta(q_0\pm\Wk{-}\mp\Wp{-})$ & \makecell[c]{ ~ \\ $  -|\qvec| \le q_0 \le |\qvec|  $ \\ ~ } \\	
\hline
\end{tabular}
\label{tab.kin.l}~.
\end{eqnarray}

In Eq.~\eqref{impi.1}, when summed over the indices $r$ and $s$, the obtained resultant kinematic regions for the Unitary and Landau cuts come out to be 
\begin{eqnarray}
	\begin{tabular}{|c|c|}
		\hline 
		Cuts & \makecell[c]{ ~ \\ Kinematic Regions \\ ~ } \\
		\hline \hline
		Unitary-I & \makecell[c]{ ~ \\ $ 2M \le q_0 < \infty$ for $|\qvec| < 2\mu_5$ \\
									$ \sqrt{(|\qvec|-2\mu_5)^2+4M^2} \le q_0 < \infty$  for $|\qvec| \ge 2\mu_5$\\ ~ }\\
		\hline
		Unitary-II & \makecell[c]{ ~ \\ $ -\infty < q_0 \le -2M $ for $|\qvec| < 2\mu_5$ \\
			$ -\infty < q_0 \le -\sqrt{(|\qvec|-2\mu_5)^2+4M^2} $  for $|\qvec| \ge 2\mu_5$\\ ~ }\\
		\hline
		Landau-I \& Landau-II & \makecell[c]{ ~ \\ $ -|\qvec|-2\mu_5 \le q_0 \le |\qvec|+2\mu_5 $ \\ ~ }\\
		\hline
	\end{tabular}
	\label{tab.range.total}~.
\end{eqnarray}

\begin{figure}[h]
	\begin{center}
		\includegraphics[angle=0,scale=0.5]{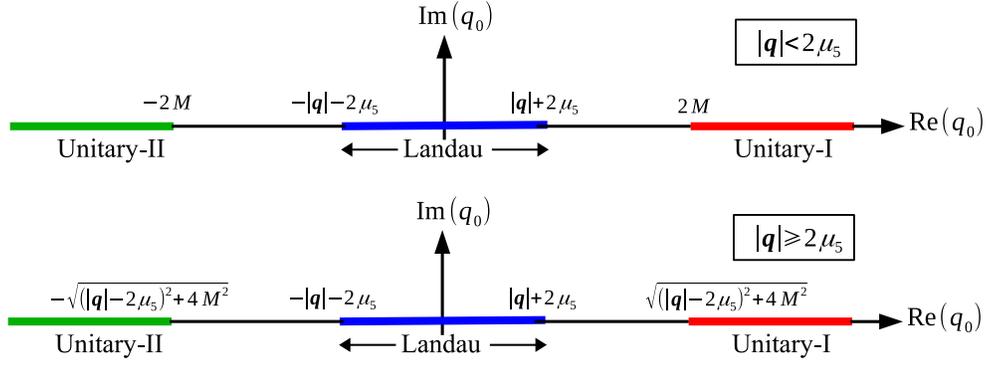} 
	\end{center}
	\caption{(Color Online) The branch cuts of the self energy in the complex $q_0$ plane for a given $|\qvec|$ when (upper panel)  $|\qvec|<2\mu_5$ (lower panel)  $|\qvec|\ge2\mu_5$.}
	\label{fig.analytic}
\end{figure}

The analytic structure of $\IM\Pi^\munu_{11} (q_0,\qvec)$ has been depicted in Fig.~\ref{fig.analytic}. If we restrict ourselves to the physical time-like kinematic domains defined in terms of $q_0>0$ and $q_0>|\qvec|$, then from Fig.~\ref{fig.analytic}, we see that along with the Unitary-I cut, some portion of the Landau cuts $|\qvec| < q_0 < 2\mu_5$ also contribute which is purely a finite CCP effect. Physically it means, a real quark/antiquark with helicity $r$ in the medium can absorbs/emits a time-like virtual photon having positive energy to become a quark/antiquark with helicity $-r$. Additionally, the threshold of the Unitary cuts strongly depend on $\mu_5$ and $M$. As a consequence, for high enough $\mu_5 \ge |\qvec|/2$, a positive energetic photon having squared invariant mass $q^2 \ge 4(M^2-\mu_5^2)$ can decay into a real quark-antiquark pair, which is is less than the usual threshold of pair production $q^2 \ge 4M^2$; the photon may even be space-like. Also, for sufficiently high $\mu_5$, the forbidden gap between the Landau and Unitary cuts will become zero irrespective of the value of $M$ which is also a purely finite CCP effect.

\section{ELECTROMAGNETIC SPECTRAL FUNCTION AND ITS LORENTZ STRUCTURE} \label{sec.lorentz}
The electromagnetic spectral function $\rho^\munu$ due to the quark loop is defined as~\cite{Mallik:2016anp,Ghosh:2018xhh}
\begin{eqnarray}
\rho^\munu (q_0,\qvec) = \frac{1}{4\pi\alpha} \sign{q_0} \tanh\FB{\frac{q_0}{2T}}\IM\Pi^\munu_{11}(q_0,\qvec)
\label{spectral.1}
\end{eqnarray}
where $\alpha = \frac{1}{137}$ is the fine structure constant. Note that, the spectral function satisfies the transversality condition $q_\mu\rho^\munu(q) = q_\nu\rho^\munu(q) = 0$ which follows from the conservation of the vector current in Eq.~\eqref{current}: $\del^\mu J_\mu^\text{em}(x) = 0$. It is useful to decompose the spectral function in suitable tensor basis constructed from the available vectors and tensors. The choice of the tensor basis is not unique, and in this work, we use the following tensor basis from Ref.~\cite{Nieves:1988qz} to decompose the spectral function
\begin{eqnarray}
\rho^\munu (q_0,\qvec) = \rho_\tL P_\tL^\munu + \rho_\tT^+ P_\tT^{+\munu} + \rho_\tT^- P_\tT^{-\munu} 
\label{spectral.2}
\end{eqnarray}
where, the projection tensors are 
\begin{gather}
P_\tL^\munu = \frac{\utilde^\mu\utilde^\nu}{\utilde^2}, \label{PL}\\
P_\tT^{\pm\munu} = \frac{1}{2} \FB{ g^\munu - \frac{q^\mu q^\nu}{q^2} - \frac{\utilde^\mu\utilde^\nu}{\utilde^2} 
	\pm \frac{1}{\sqrt{q^2\utilde^2}}\varepsilon^{\munu\alphabeta}\utilde_\alpha q_\beta } \label{PT}
\end{gather}
in which $\utilde^\mu = u^\mu - \dfrac{(q\cdot u)}{q^2} q^\mu$ is a vector orthogonal to $q^\mu$ and $u^\mu$ is the medium four-velocity vector. In the local rest frame (LRF) of the medium $u^\mu_\text{LRF} \equiv (1,{\bm 0})$. The projection tensors in Eq.~\eqref{PL} and \eqref{PT} form orthonormal basis which can be realized from the following multiplication tables:
\begingroup
\renewcommand*{\arraystretch}{2}
\setlength{\tabcolsep}{10pt} 
\begin{eqnarray}
	\begin{tabular}{|c||c|c|c|}
		\hline
		~ & $ P_\tL^{\mu\alpha} $ & $ P_\tT^{+\mu\alpha} $ & $ P_\tT^{-\mu\alpha} $ \\
		\hline \hline
		 $g_\alphabeta  P_\tL^{\beta\nu} $ & $ P_\tL^\munu $ & $ 0 $ & $ 0 $ \\
		 \hline
		 $ g_\alphabeta  P_\tT^{+\beta\nu} $ & $ 0 $ & $ P_\tT^{+\munu} $ & $ 0 $ \\
		 \hline
		 $ g_\alphabeta  P_\tT^{-\beta\nu} $  & $ 0 $ & $ 0 $ & $ P_\tT^{-\munu} $ \\
		 \hline
	\end{tabular}~~~~~~~,
\hspace{0.8cm}
	\begin{tabular}{|c||c|c|c|}
		\hline
		~ & $ g_\munu P_\tL^{\mu\alpha} $ & $ g_\munu P_\tT^{+\mu\alpha} $ & $ g_\munu P_\tT^{-\mu\alpha} $ \\
		\hline \hline 
		$g_\alphabeta  P_\tL^{\beta\nu} $ & $ 1 $ & 0 & $ 0 $ \\
		\hline
		$ g_\alphabeta  P_\tT^{+\beta\nu} $ & $ 0 $ & $ 1 $ & $ 0 $ \\
		\hline
		$ g_\alphabeta  P_\tT^{-\beta\nu} $  & $ 0 $ & $ 0 $ & $ 1 $ \\
		\hline
	\end{tabular}~.
	\label{tab.multiplication}
\end{eqnarray}
\endgroup

Using Eq.~\eqref{tab.multiplication}, it is easy to extract the form factors $\rho_\tL $ and $\rho_\tT^\pm$ from Eq.~\eqref{spectral.2} as
\begin{gather}
\rho_\tL = P_{\tL\numu} \rho^\munu = \frac{u_\mu u_\nu \rho^\munu}{\utilde^2}~, \\
\rho_\tT^\pm = P_{\tT\numu}^\pm \rho^\munu = \frac{1}{2} \FB{ \rho^\mu_{~\mu} -\rho_\tL \pm \frac{1}{\sqrt{q^2\utilde^2}} \varepsilon_{\munu\alphabeta}u^\alpha q^\beta \rho^\munu }. 
\end{gather}

In Eq.~\eqref{spectral.2}, ``~$\tL$~'' corresponds to the longitudinal mode, and, ``~$\tT^\pm$~'' refers to the transverse modes of photon propagation~\cite{Nieves:1988qz}. 
To see this, we first note that, in LRF of the medium, the various components of the projection tensors in Eq.~\eqref{PL} and \eqref{PT} reduces to:
\begin{eqnarray}
P_{\tL}^{00} &=& -\frac{\qvec^2}{q^2} ~~,~~ P_{\tL}^{0i} = P_{\tL}^{i0} = -\frac{q^0q^i}{q^2} ~~,~~ P_{\tL}^{ij} = -\frac{q_0^2}{q^2}\frac{q^iq^j}{\qvec^2} \\
P_{\tT}^{\pm 00} &=& 0 ~~,~~ P_{\tT}^{\pm 0i} = P_{\tT}^{\pm i0} = 0 ~~,~~ P_{\tT}^{\pm ij} = \frac{1}{2}\FB{-\delta^{ij} + \frac{q^iq^j}{\qvec^2} \mp i \varepsilon^{ijk}\frac{q_k}{|\qvec|}}.
\label{PT.LRF}
\end{eqnarray}
Thus in LRF, the spatial components of $P_\tL^\munu$ satisfy the relation
\begin{eqnarray}
\FB{\delta_{ij}-\frac{q_iq_j}{\qvec^2}} P_{\tL}^{jk} = 0
\end{eqnarray}
which implies that the mode ``~$\tL$~'' correspond to the longitudinal mode of the photon propagation. On the other hand, in LRF, the spatial components of $P_{\tT}^{\pm\munu}$ satisfy the orthogonality relation
\begin{eqnarray}
q^i P_{\tT ij}^\pm = 0
\end{eqnarray}
implying that the modes ``~$\tT^\pm$~'' refer to the transverse modes of the propagating photon. Moreover, if we consider the photon three-momentum $\qvec = |\qvec|\hat{\bm{z}}$ along $\hat{\bm{z}}$ direction, then we have from Eq.~\eqref{PT.LRF} that
\begin{eqnarray}
P_{\tT}^{+\munu} = - \epsilon_R^\mu \epsilon_R^{\nu*} ~~~,~~ P_{\tT}^{-\munu} = - \epsilon_L^\mu \epsilon_L^{\nu*}
\end{eqnarray}
where, 
\begin{eqnarray}
\epsilon_R^\mu = \frac{1}{\sqrt{2}}(0,1,i,0) ~~\text{and,~~} \epsilon_L^\mu = \frac{1}{\sqrt{2}}(0,1,-i,0)
\end{eqnarray}
denote the right and left circular polarization vectors. Therefore, the real poles of the propagators corresponding to  ``~$\tT^h\,(h=+,-)$~''  describe the existence of two different transverse modes identified by their circular polarization $h$. Note that at finite $T$ and $\mu$ with $\mu_5=0$, one usually has one longitudinal mode (L), and, two degenerate transverse modes (T). In presence of CCP, the longitudinal mode is unaffected, where as degeneracy of the transverse mode is lifted and we get three distinct modes of photon propagation.

\section{NUMERICAL RESULTS \& DISCUSSIONS}\label{sec.results}
\begin{figure}[h]
	\begin{center}
		\includegraphics[angle=-90,scale=0.35]{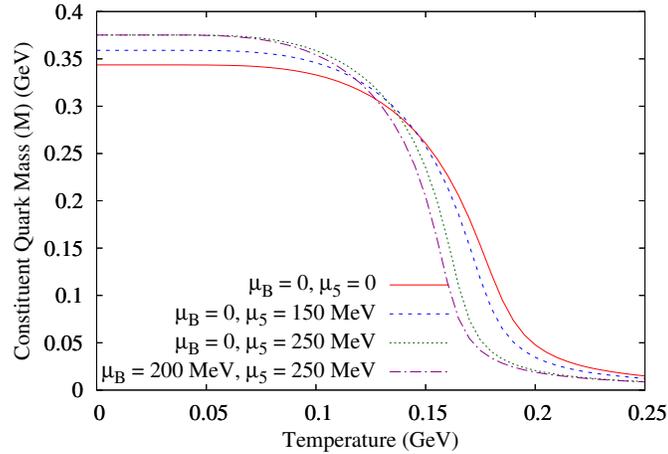} 
	\end{center}
	\caption{(Color Online) Constituent quark as a function of temperature for different values of $\mu_B$ and $\mu_5$.}
	\label{fig.M}
\end{figure}
\begin{figure}[h]
	\begin{center}
		\includegraphics[angle=-90,scale=0.70]{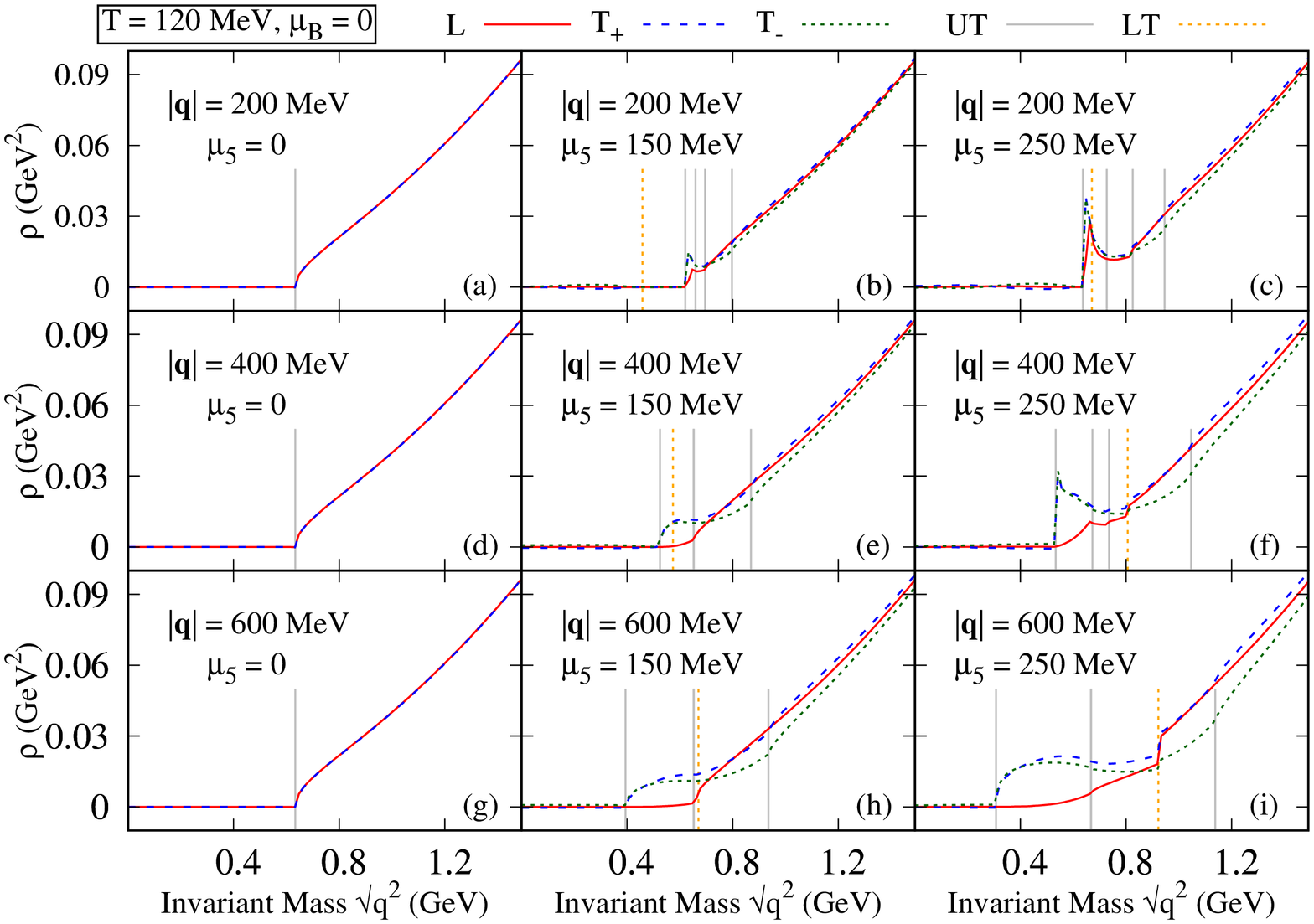}
	\end{center}
	\caption{(Color Online) Electromagnetic spectral function $\rho$ in longitudinal (L) and transverse ($\tT_\pm$) modes as a function of invariant mass $\sqrt{q^2}$ for different values of $\mu_5$ and $|\qvec|$ at $ T=120 $ MeV and $\mu_B=0$. ``UT'' and ``LT'' respectively corresponds to the Unitary and Landau cut thresholds situated in the physical time-like region ($q^0>0$ and $q^2>0$).}	
	\label{fig.spectra.1}
\end{figure}
\begin{figure}[h]
	\begin{center}
		\includegraphics[angle=-90,scale=0.70]{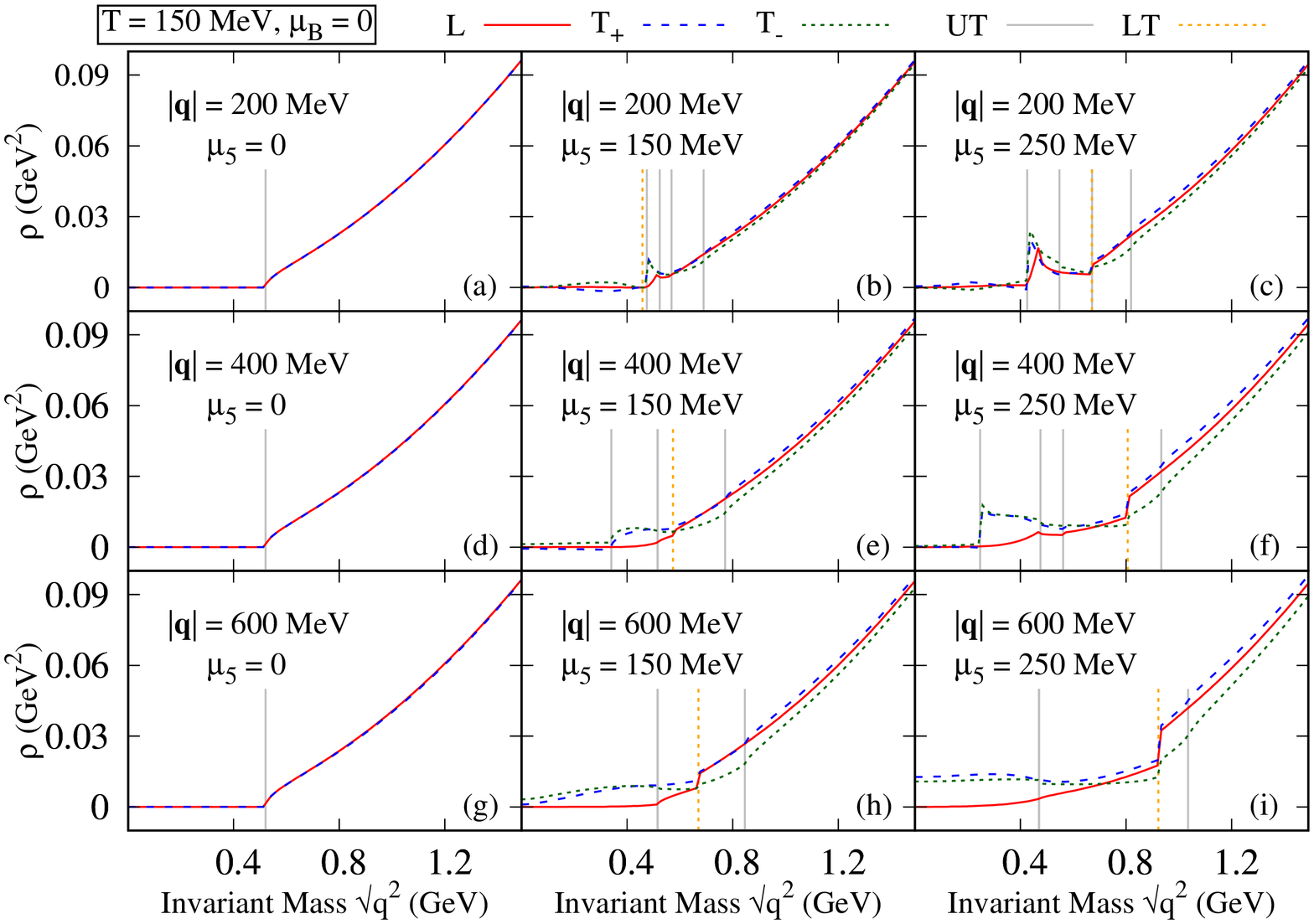}
	\end{center}
	\caption{(Color Online) Electromagnetic spectral function $\rho$ in longitudinal (L) and transverse ($\tT_\pm$) modes as a function of invariant mass $\sqrt{q^2}$ for different values of $\mu_5$ and $|\qvec|$ at $ T=150 $ MeV and $\mu_B=0$. ``UT'' and ``LT'' respectively corresponds to the Unitary and Landau cut thresholds situated in the physical time-like region ($q^0>0$ and $q^2>0$).}	
	\label{fig.spectra.2}
\end{figure}
\begin{figure}[h]
	\begin{center}
		\includegraphics[angle=-90,scale=0.70]{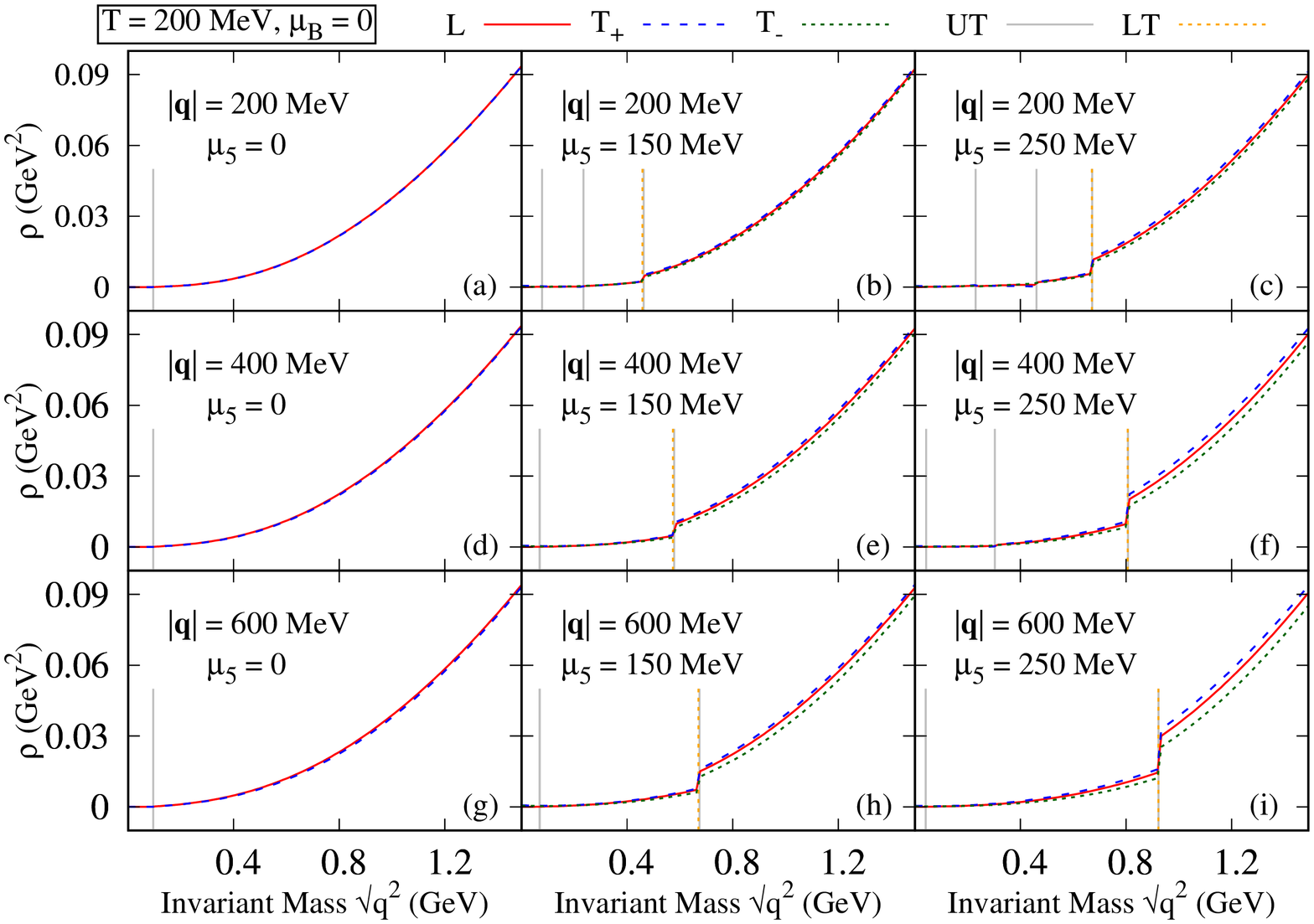}
	\end{center}
	\caption{(Color Online) Electromagnetic spectral function $\rho$ in longitudinal (L) and transverse ($\tT_\pm$) modes as a function of invariant mass $\sqrt{q^2}$ for different values of $\mu_5$ and $|\qvec|$ at $ T=200 $ MeV and $\mu_B=0$. ``UT'' and ``LT'' respectively corresponds to the Unitary and Landau cut thresholds situated in the physical time-like region ($q^0>0$ and $q^2>0$).}	
	\label{fig.spectra.3}
\end{figure}
%
%
\begin{figure}[h]
	\begin{center}
		\includegraphics[angle=-90,scale=0.70]{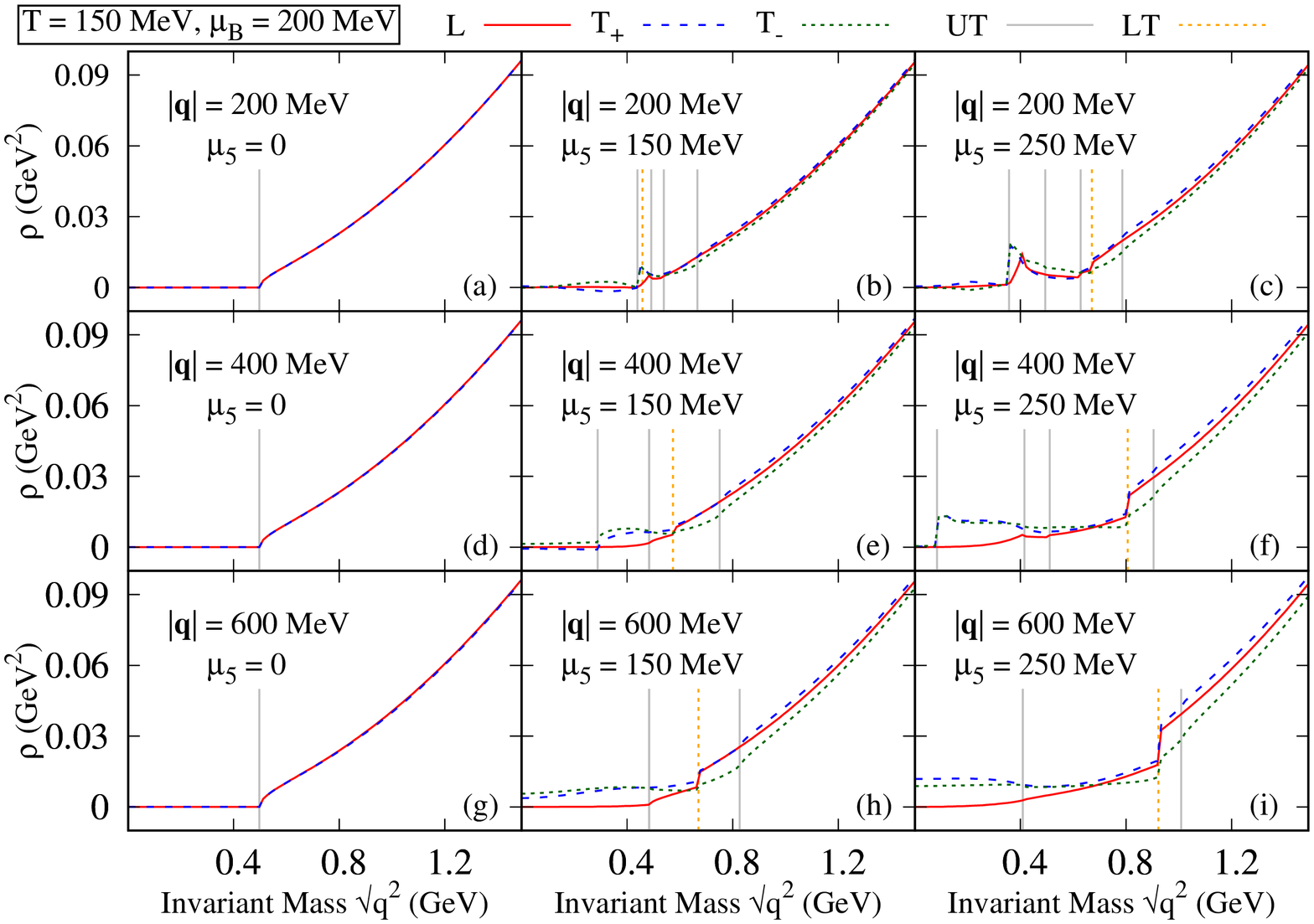}
	\end{center}
	\caption{(Color Online) Electromagnetic spectral function $\rho$ in longitudinal (L) and transverse ($\tT_\pm$) modes as a function of invariant mass $\sqrt{q^2}$ for different values of $\mu_5$ and $|\qvec|$ at $T=150$ MeV and $\mu_B=200$ MeV. ``UT'' and ``LT'' respectively corresponds to the Unitary and Landau cut thresholds situated in the physical time-like region ($q^0>0$ and $q^2>0$).}	
	\label{fig.spectra.5}
\end{figure}
\begin{figure}[h]
	\begin{center}
		\includegraphics[angle=-90,scale=0.70]{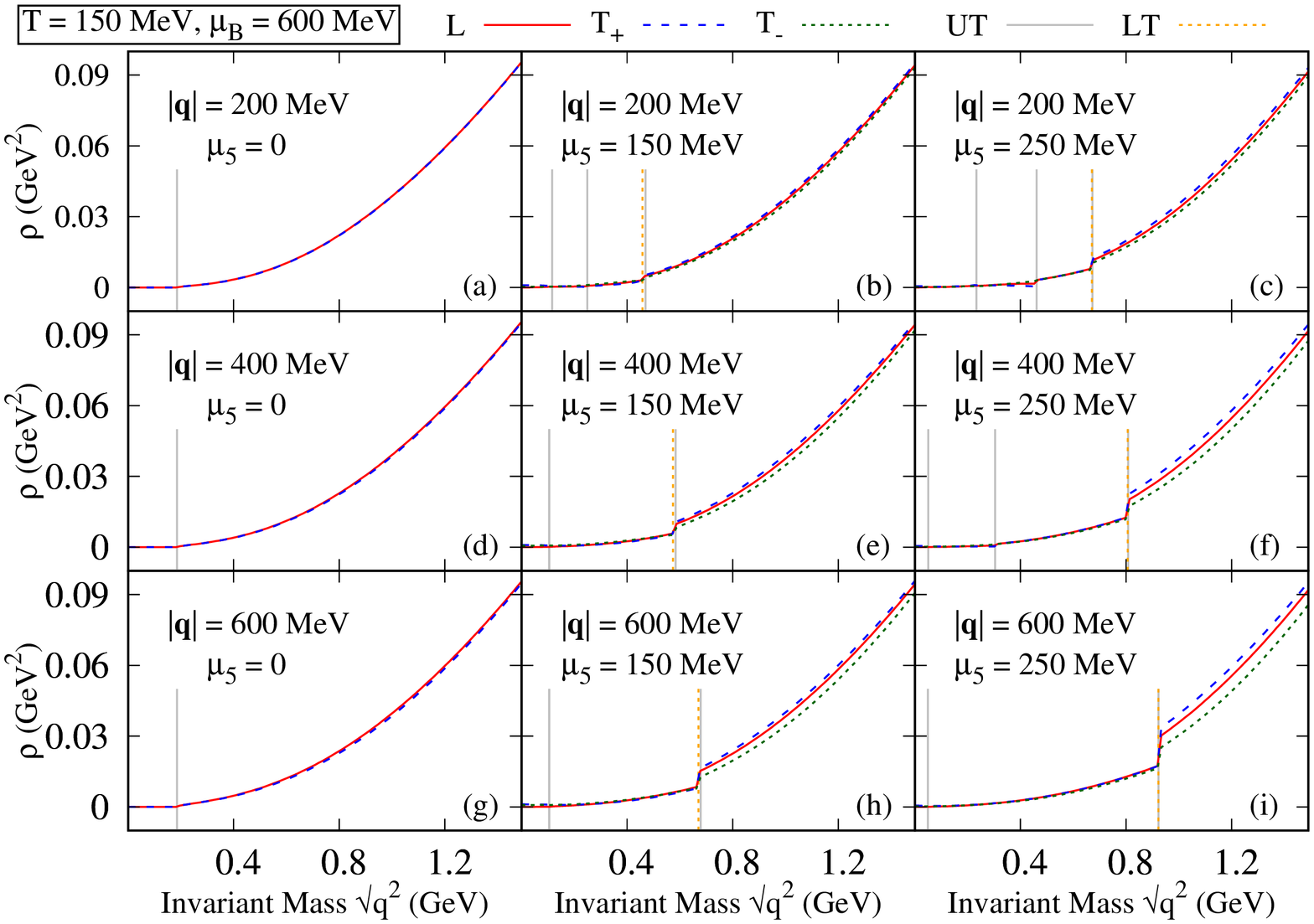}
	\end{center}
	\caption{(Color Online) Electromagnetic spectral function $\rho$ in longitudinal (L) and transverse ($\tT_\pm$) modes as a function of invariant mass $\sqrt{q^2}$ for different values of $\mu_5$ and $|\qvec|$ at $T=150$ MeV and $\mu_B=600$ MeV. ``UT'' and ``LT'' respectively corresponds to the Unitary and Landau cut thresholds situated in the physical time-like region ($q^0>0$ and $q^2>0$).}	
	\label{fig.spectra.6}
\end{figure}
Let us first specify the values of the parameters of the NJL model that are used for the numerical results. We use current quark mass $m=5.6$ MeV, smooth three-momentum cutoff $\Lambda=568.69$ MeV and NJL coupling $G=5.742$ GeV$^{-2}$ which reproduce the experimental/phenomenological vacuum values of quark-condensate, pion decay constant and pion mass. We now show the variation of the constituent quark mass $M$ as a function of temperature for different values of BCP and CCP in Fig.~\ref{fig.M} as obtained by solving the gap equation~\eqref{gap.final}. We notice that, $M$ is large ($\sim 300$ MeV) in the low temperature region owing to the spontaneous breaking of the chiral symmetry yielding large values of the quark condensate. With the increase in temperature, $M$ first remain constant upto a certain value of temperature after which $M$ suddenly decreases corresponding to the pseudo-chiral phase transition. In the high temperature limit, $M$ approaches to the current quark mass $m$ irrespective of the values of other parameters like BCP and CCP. With the increase in BCP, the transition temperature is found to decrease thus mimicking a typical QCD phase diagram.

By comparing the red, blue and green curves of Fig.~\ref{fig.M}, we  observe that, with the increase in CCP, the constituent quark mass $M$ increases in the low temperature region; however the sudden change of $M$ occurs at a relatively smaller temperature value indicating a decrease in the transition temperature with the increase in CCP. Hence, CCP has the tendency to make the chiral condensate stronger at low temperature $T\simeq0$ (which is similar to the magnetic catalysis) and may be called ``chiral catalysis'' where the chiral imbalance catalyzes the dynamical symmetry breaking. On the other hand, at large temperature, an opposite effect is noticed where the presence of non-zero CCP weakens the chiral condensate and chiral symmetry is restored at a lower temperature as compared to the vanishing CCP case. This may be termed as ``inverse chiral catalysis'' (similar to the inverse magnetic catalysis) where an chiral imbalance opposes the dynamical symmetry breaking.

Let us now switch to the numerical results of the electromagnetic spectral functions for which we restrict ourselves to physical time-like kinematic region defined in terms of $q^0>0$ and $q^2>0$. In Fig.~\ref{fig.spectra.1}, we have depicted the variation of the electromagnetic spectral function $\rho$ in longitudinal (L) and transverse ($\tT_\pm$) modes as a function of invariant mass $\sqrt{q^2}$ for different values of $\mu_5$ and $|\qvec|$ at $ T=120 $ MeV and $\mu_B=0$. In Fig.~\ref{fig.spectra.1}, we have also shown the positions of the Unitary and Landau cut thresholds which appeared in the physical kinematic region by grey and orange horizontal lines respectively. Noticing Sub-Figs.~\ref{fig.spectra.1}(a), (d) and (g), we find that for vanishing CCP, the longitudinal and the degenerate transverse spectral functions have similar magnitudes and their difference is not visible with the range of the plot. Moreover, at $\mu_5=0$, we have only the Unitary-I cut threshold $\sqrt{q^2}>2M$ in the physical region which does not depend on $|\qvec|$.

Observing the non-zero CCP graphs in Fig.~\ref{fig.spectra.1}, we find more than one Unitary-I cut thresholds (corresponding to sub-cuts $\{r,s\} = \{+,+\}, \{+,-\}, \{-,+\}$ and $ \{-,-\}$) and one Landau cut threshold (corresponding to sub-cuts $\{r,s\} = \{+,-\}$ or $\{-,+\}$) in the physical region as can be understood from Eqs.~\eqref{tab.kin.u} and \eqref{tab.kin.l}. In Sub-Figs.~\ref{fig.spectra.1}(e), (h) and (i), we see three Unitary cut thresholds instead of four since the sub-cuts $\{r,s\} = \{+,-\}$ and $\{-,+\}$ have identical thresholds as per Eq.~\eqref{tab.kin.u}. In Sub-Fig.~\ref{fig.spectra.1}(b), it is found that, there exists a kinematic gap between the Landau cut and the Unitary cut where the spectral function is zero. In the rest of the non-zero CCP graphs, the Landau cut threshold appears within the Unitary owing to the overlap of the cuts, yielding a continuous spectrum of electromagnetic spectral function over the whole range of invariant mass. We emphasize that, the generation of the continuous spectrum of spectral function is purely a finite CCP effect.

The finite CCP graphs in Fig.~\ref{fig.spectra.1} is seen to have non-monotonic dependence on the invariant mass. Since the contribution of the different sub-cuts to the spectral function are not at par in magnitude, we see sharp changes of the spectral function values at each sub-cut thresholds. From Fig.~\ref{fig.spectra.1}, it is also interesting to note that, at finite value of CCP, the degeneracy of the transverse modes are lifted and we get three distinct modes of the spectral function. Moreover, unlike the zero-CCP case, the distinction among the modes are clearly visible in the finite CCP plots. The differences among the longitudinal (L) and two non-degenerate transverse ($\tT_\pm$) modes increase as we increase $|\qvec|$ and $\mu_5$.

Next in Fig.~\ref{fig.spectra.2}, we have shown the variation of the spectral function $\rho$ in different modes as a function of invariant mass for different values of $\mu_5$ and $|\qvec|$ at $ T=150 $ MeV and $\mu_B=0$. As compared to the $ T=120 $ MeV case (in Fig.~\ref{fig.spectra.1}), the constituent quark mass has substantially decreased at $ T=150 $ MeV. Thus, the Unitary cut thresholds have moved towards lower values of the invariant mass in Fig.~\ref{fig.spectra.2} compared to that of Fig.~\ref{fig.spectra.1}. On the other hand, the Landau cut threshold has weaker dependence on $M$ (due to the relative signs of the terms in middle rows of Eq.~\eqref{tab.kin.l} as compared to Eq.~\eqref{tab.kin.u}) and has strong dependence on the other parameters (like $|\qvec|$ and $\mu_5$). Hence comparing Fig.~\ref{fig.spectra.2} and Fig.~\ref{fig.spectra.1}, we notice that unlike the Unitary-cut thresholds, the Landau cut threshold does not move significantly over the invariant mass axes.

As before, in Sub-Figs.~\ref{fig.spectra.2}(e), (h) and (i), we found that, the Unitary sub-cuts $\{r,s\} = \{+,-\}$ and $\{-,+\}$ have identical thresholds. More interestingly, in Sub-Figs.~\ref{fig.spectra.2}(h) and (i), the Unitary sub-cut $\{r,s\} = \{-,-\}$ is absent as it has started from the space-like kinematic region ($q^2<0$) as can be understood from last row of Eq.~\eqref{tab.kin.u}. The qualitative nature of the graphs of Fig.~\ref{fig.spectra.2} is found to be similar to Fig.~\ref{fig.spectra.1}.

Next in Fig.~\ref{fig.spectra.3}, we have depicted the variation of the spectral function $\rho$ in longitudinal (L) and transverse ($\tT_\pm$) modes as a function of invariant mass for different values of $\mu_5$ and $|\qvec|$ at $ T=200 $ MeV and $\mu_B=0$ which is the region in NJL phase diagram with the partially restored chiral symmetry. In this case, the constituent quark mass $M\simeq m$, so that, the Unitary cut thresholds have further moved towards lower values of the invariant mass as compared to that of Figs.~\ref{fig.spectra.1} and \ref{fig.spectra.2}. As expected from Eq.~\eqref{tab.kin.l}, the Landau cut threshold does not move appreciably along the invariant mass axes. In Sub-Figs.~\ref{fig.spectra.3}(e), (h) and (i), we again found that, the Unitary sub-cuts $\{r,s\} = \{+,-\}$ and $\{-,+\}$ have identical thresholds. Additionally, in Sub-Figs.~\ref{fig.spectra.2}(b) (c), (e), (f), (h) and (i), the Unitary sub-cut $\{r,s\} = \{-,-\}$ is found to be absent as it has started from the space-like kinematic region ($q^2<0$) due to the small value of $M\simeq m$. 

Until now, we have considered the BCP to be zero. Let us now show the numerical results of the spectral functions for finite $\mu_B$ in Figs.~\ref{fig.spectra.5} and \ref{fig.spectra.6}. In Fig.~\ref{fig.spectra.5}, we have shown the variation of the spectral function $\rho$ in different modes as a function of invariant mass for different values of $\mu_5$ and $|\qvec|$ at $ T=150 $ MeV and $\mu_B=200$ MeV. Comparing non-zero BCP graphs of Fig.~\ref{fig.spectra.5} with the corresponding $\mu_B=0$ graphs of Fig.~\ref{fig.spectra.2}, we find that the Unitary-cut thresholds have moved towards the lower invariant mass due to the decrease in $M$. Both the qualitative and quantitative nature of the curves of Fig.~\ref{fig.spectra.5} is seen to be similar to the Fig.~\ref{fig.spectra.2}.

Finally, in Fig.~\ref{fig.spectra.6}, we have plotted the variation of the spectral function $\rho$ in longitudinal (L) and transverse ($\tT_\pm$) modes as a function of invariant mass for different values of $\mu_5$ and $|\qvec|$ at $ T=150 $ MeV and $\mu_B=600$ MeV which is again the region in NJL phase diagram with the partially restored chiral symmetry. In this case, the constituent quark mass $M\simeq m$, so that, the nature of the all the graphs become similar (both qualitatively and quantitatively) to the corresponding $ T=200 $ MeV 
graphs of Fig.~\ref{fig.spectra.3} 
.


\section{SUMMARY \& CONCLUSION}\label{sec.summary}
In summary, we have studied the electromagnetic spectral function of hot and dense quark matter with chiral imbalance. This is done by evaluating the imaginary part of the one-loop photon self energy at finite temperature, BCP and CCP employing the real time formalism of finite temperature field theory. The effect of ``strong'' interaction have been incorporated by means of a temperature, BCP and CCP dependent constituent quark mass $M=M(T,\mu,\mu_5)$ obtained from a two-flavour NJL model. Incorporation of a CCP in the NJL model was found to have interesting consequences on the constituent mass of quarks obtained from the gap equation. It was found that the chiral condensate tends to get stronger at low temperature while the opposite is observed at high values of temperature. These could be termed as ``chiral catalysis'' and ``inverse chiral catalysis'' respectively.

A study of the analytic structure of the electromagnetic spectral function in the complex energy plane revealed a rich structure with multiple Landau and Unitary type discontinuities. Again, three distinct modes of the spectral function could be observed on account of the lifting of degeneracy of the transverse modes in the presence of CCP. Most interestingly, a continuous spectrum is obtained for the electromagnetic spectral function and this can be attributed purely to the presence of a CCP. This in turn could have interesting consequences on the spectra of electromagnetic probes from chiral imbalanced matter.

\section*{Acknowledgments}
S.G. is funded by the Department of Higher Education, Government of West Bengal, India. N.C., S.S. and P.R. are funded by the Department of Atomic Energy (DAE), Government of India. 


\appendix
\section{A NOTE ON INTRODUCING FERMION CHEMICAL POTENTIAL IN RTF} \label{app.mu}
Let us start with the simple Dirac Lagrangian density
	\begin{eqnarray}
		\scrL = \Psibar\FB{i\gamma^\mu \del_\mu -M}\Psi
		\label{eq.lag.vac}
	\end{eqnarray}
yielding the following equation of motion 
\begin{eqnarray}
\FB{i\gamma^\mu \del_\mu -M}\Psi(x) = 0,
\label{eq.eom}
\end{eqnarray}
where $\Psi$ denotes the fermion field. Now, in finite temperature field theory, the temperature $T=1/\beta$ and the chemical potential $\mu$ corresponding to the conserved charge $Q = \int d^3x J^0(x)$ can enter through the density operator~\cite{Kapusta:2006pm,Bellac:2011kqa} 
\begin{eqnarray}
\rho^\text{GC} = e^{-\beta(H-\mu Q)} \label{eq.density.op.grand}
\end{eqnarray}
of the grand canonical (GC) ensemble where, $H$ is the Hamiltonian and $J^\mu(x)$ is the conserved current corresponding to $Q$. Using Eq.~\eqref{eq.density.op.grand}, the coordinate space thermo-dense fermion propagator in RTF is defined as the thermal average of two-point correlation function in GC ensemble as
\begin{eqnarray}
S^\text{GC}(x,x') = i \ensembleaverage{\mcTc \Psi(\tau,\xvec)\Psibar(\tau',\xvec')}^\text{GC} = \frac{ i\Tr \TB{ \rho^\text{GC} \mcTc \Psi(\tau,\xvec)\Psibar(\tau',\xvec')} }{\Tr \TB{ \rho^\text{GC} } }
\label{eq.Sx}
\end{eqnarray}
where, $\mcTc$ denotes the time ordering with respect to a contour $C$ in the complex time ($\tau$) plane (in RTF, the choice of $C$ is not unique). We also note that, $S^\text{GC}(x,x')$ is the Green's function of Eq.~\eqref{eq.eom} as~\cite{Mallik:2016anp}
\begin{eqnarray}
\FB{i\gamma^\mu \del_\mu -M}S^\text{GC}(x,x') = -\delta_C(\tau-\tau')\delta^3(\xvec-\xvec').
\label{eq.green}
\end{eqnarray}

On the other hand, an alternative way to incorporate $\mu\ne0$, is to replace the Hamiltonian~\cite{Bellac:2011kqa,Kapusta:2006pm} 
\begin{eqnarray}
H \to \widetilde{H} = \FB{H -\mu Q} = H - \mu \int d^3x J^0(x)
\label{eq.H}
\end{eqnarray}
so that the Lagrangian density modifies to 
\begin{eqnarray}
\scrL \to \widetilde{\scrL} = \FB{\scrL + \mu J^0}, 
\label{eq.L}
\end{eqnarray}
and then use the canonical density operator $\rho^\text{C} = e^{-\beta H}$ for ensemble average. In this case, Eq.~\eqref{eq.Sx} modifies to
\begin{eqnarray}
S^\text{C}(x,x') = i \ensembleaverage{\mcTc \ut{\Psi}(\tau,\xvec)\ut{\Psibar}(\tau',\xvec')}^\text{C} = \frac{ i\Tr \TB{ \rho^\text{C} \mcTc \ut{\Psi}(\tau,\xvec)\ut{\Psibar}(\tau',\xvec')} }{\Tr \TB{ \rho^\text{C} } }
\label{eq.Sx.Can}
\end{eqnarray}
where $\ut{\Psi}$ satisfies the modified equation of motion that follows from $\widetilde{\scrL}$ in Eq.~\eqref{eq.L}.

Now, as an example, the Lagrangian in Eq.~\eqref{eq.lag} is invariant under the global phase transformation $\Psi \to e^{i\theta_V}\Psi$ owing to $U_V(1)$ global gauge invariance. From the corresponding Noether's current $J^\mu = \Psibar\gamma^\mu\Psi$, we have the conserved charge $Q  = \int d^3x \Psibar(x)\gamma^0 \Psi(x) = \int d^3x \Psi^\dagger(x) \Psi(x)$. Thus the modified Lagrangian of Eq.~\eqref{eq.L} becomes
\begin{eqnarray}
\widetilde{\scrL} = \ut{\Psibar}\FB{i\gamma^\mu \del_\mu + \mu \gamma^0 -M}\ut{\Psi}
\label{eq.Ltil}
\end{eqnarray}
which yields the modified equation of motion as 
\begin{eqnarray}
\FB{i\gamma^\mu \del_\mu + \mu \gamma^0 -M}\ut{\Psi} = 0.
\end{eqnarray}
Therefore, the Green's function $S^\text{C}(x,x')$ will satisfy
\begin{eqnarray}
\FB{i\gamma^\mu \del_\mu + \mu\gamma^0 -M}S^\text{C}(x,x') = -\delta_C(\tau-\tau')\delta^3(\xvec-\xvec').
\end{eqnarray}

It has been shown in Refs.~\cite{Niegawa:2002wj,Bellac:2011kqa} that, the two different forms of the propagator namely $S^\text{GC}(x,x')$ and $S^\text{C}(x,x')$ are related via the relation
\begin{eqnarray}
S^\text{C}(x,x') = e^{i\mu(\tau-\tau')}S^\text{GC}(x,x'),
\end{eqnarray}
or in momentum space (for example the $11$-component) via
\begin{eqnarray}
S^\text{C}_{11}(p^0,\pvec) = S^\text{GC}_{11}(p^0+\mu,\pvec).
\end{eqnarray}
In standard perturbative calculations such as the one-loop calculations performed in this work, using any of the two different forms of the propagator will lead to the same final result~\cite{Niegawa:2002wj}.

\section{} \label{app.domain}
We tabulate respectively the range (or co-domain) of the functions $U^{rs}(|\kvec|,x; M,|\vec{q}|,\mu_5)$ and $L^{rs}(|\kvec|,x; M,|\vec{q}|,\mu_5)$ (defined in Eqs.~\eqref{U} and \eqref{L}) in Eqs~\eqref{tab.range.u} and \eqref{tab.range.l} below:
%
%
\begin{eqnarray}
	\begin{tabular}{|c|c|}
		\hline 
		$ (r,s) $ & \makecell[c]{ ~ \\ $\ran(U^{rs})$ \\ ~ } \\
		\hline \hline
		(+,+) & \makecell[c]{ ~ \\ $\Big[ \sqrt{(|\qvec|+2\mu_5)^2+4M^2} , \infty\Big)$ \\ ~ }\\
		\hline
		(+,-) & \makecell[c]{ ~ \\ $\Big[\sqrt{\mu_5^2+M^2} + \sqrt{(|\qvec|-\mu_5)^2+M^2} , \infty \Big)$ for $ |\qvec| < 2\mu_5 $, \\
							$\Big[\sqrt{\qvec^2+4M^2} , \infty\Big)$  for $ |\qvec| \ge 2\mu_5 $ \\ ~ } \\
		\hline 
		(-,+) & \makecell[c]{~ \\ $\Big[ \frac{1}{2}\sqrt{(|\qvec|-2\mu_5)^2+4M^2} + \frac{1}{2}\sqrt{(|\qvec|+2\mu_5)^2+4M^2} , \infty \Big)$ for $ |\qvec| < \mu_5 $, \\
			$\Big[ \frac{1}{2}\sqrt{\qvec^2+4M^2} + \frac{1}{2}\sqrt{(||\qvec|-2\mu_5|+2\mu_5)^2+4M^2} , \infty\Big)$  for $ |\qvec| \ge \mu_5 $ \\ ~ } \\
		\hline
		(-,-) & \makecell[c]{ ~ \\ $\Big[ 2M , \infty \Big)$ for $ |\qvec| < 2\mu_5 $, \\
		$\Big[  \sqrt{(|\qvec|-2\mu_5)^2+4M^2} , \infty\Big)$  for $ |\qvec| \ge 2\mu_5 $ \\ ~} \\	
	\hline
	\end{tabular}
\label{tab.range.u}
\end{eqnarray}
\begin{eqnarray}
	\begin{tabular}{|c|c|}
	\hline 
	$ (r,s) $ & \makecell[c]{ ~ \\ $\ran(L^{rs})$ \\ ~ } \\
	\hline \hline
	$ (+,+) $ &  \makecell[c]{~ \\ $\Big[ -|\qvec| , |\qvec| \Big]$ \\ ~ } \\
	\hline
	$ (+,-) $ & \makecell[c]{ ~ \\ $\Big[ -|\vec{q}|-2\mu_5 , -\sqrt{\mu_5^2+M^2} + \sqrt{(|\qvec|-\mu_5)^2+M^2} \Big]$ for $ |\qvec| < 2\mu_5 $, \\
		$\Big[ -|\vec{q}|-2\mu_5 , |\vec{q}|-2\mu_5 \Big]$  for $ |\qvec| \ge 2\mu_5 $ \\ ~ } \\
	\hline 
	$ (-,+) $ & \makecell[c]{ ~ \\ $\Big[ -\frac{1}{2}\sqrt{(|\qvec|-2\mu_5)^2+4M^2} + \frac{1}{2}\sqrt{(|\qvec|+2\mu_5)^2+4M^2} , |\vec{q}|+2\mu_5 \Big]$ for $ |\qvec| < \mu_5 $, \\
		$\Big[ -\frac{1}{2}\sqrt{\qvec^2+4M^2} + \frac{1}{2}\sqrt{(|\qvec|-4\mu_5)^2+4M^2} , |\vec{q}|+2\mu_5 \Big]$  for $ \mu_5 \le |\qvec| < 2\mu_5 $, \\
		$\Big[ -|\vec{q}|+2\mu_5 , |\vec{q}|+2\mu_5 \Big]$  for $ |\qvec| \ge 2\mu_5 $ \\ ~ } \\
	\hline
	$ (-,-) $ & \makecell[c]{ ~ \\ $ \Big[ -|\qvec| , |\qvec| \Big] $ \\ ~ } \\	
	\hline
\end{tabular}
	\label{tab.range.l}
\end{eqnarray}

\bibliographystyle{apsrev4-1}
\bibliography{zDon}

\end{document}